\documentclass[twocolumn,prc,aps,showpacs,amsmath,amssymb,nofootinbib]{revtex4}
\usepackage{graphicx}

\def\im{{\rm Im}}
 \def\Pom{{ I\!\!P}}
 \def\Reg{{ I\!\!R}}
 \def\gsim{\mathrel{\rlap{\lower4pt\hbox{\hskip1pt$\sim$}}
 \raise1pt\hbox{$>$}}}

 \newcommand\la{\langle}
 \newcommand\ra{\rangle}
 \newcommand\beq{\begin{equation}}
 
 \newcommand\eeq{\end{equation}}
 \newcommand\beqn{\begin{eqnarray}}
 \newcommand\eeqn{\end{eqnarray}}
\def\mb{\,\mbox{mb}}
\def\fm{\,\mbox{fm}}
\def\GeV{\,\mbox{GeV}}

\def\lsim{\mathrel{\rlap{\lower4pt\hbox{\hskip1pt$\sim$}}
    \raise1pt\hbox{$<$}}}         
\def\gsim{\mathrel{\rlap{\lower4pt\hbox{\hskip1pt$\sim$}}
    \raise1pt\hbox{$>$}}}         

\def\Im{\,{\rm Im}\,}
\def\mb{\,\mbox{mb}}
\def\fm{\,\mbox{fm}}
\def\GeV{\,\mbox{GeV}}

\begin{document}
\date{}

\title{\bf Damping of forward neutrons in \boldmath$pp$ collisions}

\author{B.Z.~Kopeliovich$^{a,b}$}
\author{I.K.~Potashnikova$^{a}$}
\author{Ivan~Schmidt$^a$}
\author{J.~Soffer$^{c}$}

\affiliation{$^a$Departamento de F\'\i sica
y Centro de Estudios
Subat\'omicos,\\ Universidad T\'ecnica
Federico Santa Mar\'\i a, Casilla 110-V, Valpara\'\i so, Chile\\
$^b$Joint Institute for Nuclear Research, Dubna, Russia\\
$^c$Department of Physics, Temple University, Philadelphia, PA 19122-6082,
USA}

\date{\today}

\begin{abstract}

We calculate absorptive corrections to single pion exchange in the
production of leading neutrons in $pp$ collisions. Contrary to the
usual procedure of convolving the survival probability with the
cross section, we apply corrections to the spin amplitudes. The
non-flip amplitude turns out to be much more suppressed by
absorption than the spin-flip one. We identify the projectile proton
Fock state responsible for the absorptive corrections as a color
octet-octet 5-quarks configuration. Calculations within two very
different models, color-dipole light-cone description, and in
hadronic representation, lead to rather similar absorptive
corrections. We found a much stronger damping of leading neutrons
than in some of previous estimates. Correspondingly, the cross section is
considerably smaller than was measured at ISR. However, comparison
with recent measurements by the ZEUS collaboration of neutron
production in deep-inelastic scattering provides a strong motivation 
for challenging   the normalization of the ISR data. This conjecture is 
also supported by preliminary data from the NA49 experiment for
neutron production in $pp$ collisions at SPS.

\end{abstract}


\maketitle

\section{Introduction}\label{intro}

The pion is known to have a large coupling to nucleons, therefore
pion exchange is important in processes with isospin one in the
cross channel (e.g. $p+n\to n+p$). However, the pion Regge
trajectory has a low intercept $\alpha_\pi(0)\approx 0$, and this is
why it ceases to be important at high energies in binary reactions,
while other mesons, $\rho,\ a_2$, etc. take over.

\begin{figure}[htb]
\centerline{
  \scalebox{0.41}{\includegraphics{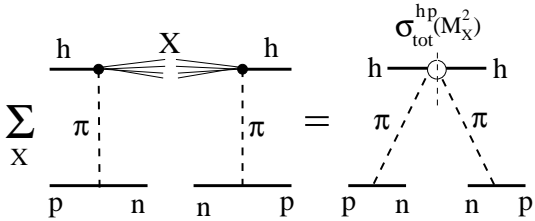}}}
\caption{\label{3r-pion} Graphical representation of the cross
section of inclusive neutron production in hadron-proton collisions,
in the fragmentation region of the proton. }
 \end{figure}

Quite a different situation occurs in inclusive reactions of leading
neutron production. Inclusive reactions in general are known to have
(approximate) Feynman scaling, and as a consequence the pion
contribution to neutron production remains nearly unchanged with
energy. This can be seen from the graphical representation of the
cross section of the inclusive reaction $h+p\to X+n$, depicted in
Fig.~\ref{3r-pion}.
 Summing up all final states $X$ at a fixed invariant mass $M_X$ and
relying on the optical theorem, one arrives at the total hadron-pion
cross section at c.m. energy $M_X$. This cross section is a slowly
varying function of $M_X$ (restricted by the Froissart bound), and
this is the source for Feynman scaling. At the same time, the
effective interval of energy squared for pion exchange is less than
$s$, which is the c.m. energy squared for $hp$ collisions. Indeed,
the effective energy squared interval $s'$ is given by the
multi-peripheral kinematics of particle production as,
 \beq
{s'\over s_0}={s\over M_X^2}\approx {1\over 1-z}\,,
\label{50}
 \eeq
 where $s_0$ is the scale factor, usually fixed at $1\GeV^2$; and
$z=p_n^+/p_p^+$ is the fraction of the proton light-cone momentum
carried by the neutron, which is close to Feynman $x_F$ at large
$z\to1$.

In fact, the pion exchange brings in a factor $(1-z)^{-2\alpha_\pi}$
($\alpha_\pi(t)$ is the pion Regge trajectory) to the cross section,
which is independent of the collision energy $s$, if $z$ is fixed.
Thus, the pion exchange contribution does not vanish with energy,
and this is in more detail the origin of the Feynman scaling. From
the point of view of dispersion relations, the smaller the
4-momentum transfer squared $t$, the closer we approach the pion
pole, and the more important is its contribution. The smallest
values of $t$ are reached in the forward direction and at $z\to1$.
The latter condition, however, leads to the dominance of other
Reggeons which have higher intercepts. Indeed, the corresponding
Regge factor $(1-z)^{-2\alpha_{\footnotesize\Reg}}$ for $\rho$ and
$a_2$ Reggeons is about $1/(1-z)$ times larger than the one for
pion. Although in general these Reggeons are suppressed by an order
of magnitude compared to the pion \cite{k2p}, they become equally
important and start taking over at $z\gsim0.9$.

Another important correction, which is the main focus of this paper,
is the effect of absorption, or initial/final state interactions.
The active projectile partons participating in the reaction, as well
as the spectator ones, can interact with the proton target or with
the recoil neutron, and initiate particle production, which usually
leads to a substantial reduction of the neutron momentum. The
probability that this does not happen, called sometimes survival
probability of a large rapidity gap, leads to a suppression of
leading neutrons produced at large $z$. There are controversies
regarding the magnitude of this suppression. Some calculations
predict quite a mild effect, of about $10\%$
\cite{boreskov1,boreskov2,3n,ap}, while others
\cite{strong1,strong2,ryskin} expect a strong reduction by about a
factor of 2. See \cite{ryskin} for a discussion of the current
controversies in data and theory, for leading neutron production.

Usually absorptive corrections are calculated in a probabilistic
way, convolving the gap survival probability with the cross
section. We found, however, that the spin amplitudes of neutron
production acquire quite different suppression factors, and one
should work with amplitudes, rather than with probabilities.

In Sect.~\ref{pion.pole} we introduce the spin amplitudes for
inclusive production of neutrons and calculate the cross section in
Born approximation of single pion exchange. Contrary to the usual
case in binary reactions, the spin non-flip term is large and rises
towards small $z$. Comparison with ISR measurements \cite{isr} shows
that the calculation overshoots somewhat the data, albeit only by
about $10\%$. Calculations also result in a substantial rise of the
cross section with energy.

In Sect.~\ref{absorptive.corrections} the absorptive corrections are
introduced. Assuming that the corrections factorize in impact
parameter space, the spin amplitudes are transformed to this
representation, and the general expression for the gap survival
amplitude is derived. We found that the main Fock component of the
incoming proton, which is responsible for the absorptive
corrections, is a 5-quark color octet-octet state. Therefore it is
not a surprise that the resulting neutron damping at which we arrive
is quite strong. In order to figure out what was missed in previous
calculations which led to a weak absorption damping, in
Sect.~\ref{sub-reggeons} we reformulated the current mechanism in
terms of Reggeon calculus.

We calculate the gap survival amplitude within two quite different
models. In Sect.~\ref{dipole} we employ the well developed
phenomenology of light-cone color dipoles fitted to photoproduction
and deep-inelastic scattering (DIS) data. We use the saturated model
for the dipole cross section, generalized recently to a partial
dipole-proton amplitude.

Another model for the survival amplitude is presented in
Sect.~\ref{hadronic}. Expanding the 5-quark Fock state over the full
set of hadronic states, we assumed that the $\pi p$ pair containing
the 5 valence quark is the dominant term. The gap survival
amplitudes of pion and proton was extracted in a model independent
way directly from data for elastic $\pi p$ and $pp$ scattering. We
found that the results of the two models, based on dipole and
hadronic representations, resulted in rather similar gap survival
amplitudes.

In Sect.~\ref{xsection} we calculate the spin non-flip and flip
contributions to the cross section, and found that the inclusive
cross section of neutron production is about twice as small as the
original result of the Born approximation. We also conclude that
absorptive corrections practically terminate the strong energy
dependence that results from the Born approximation. The ISR data
support this observation.

Although the calculated shape of $z$-distribution is improved by
absorption and corresponds to the shape of the ISR data at $q_T=0$,
the overall normalization is quite lower than in the data. In
Sect.~\ref{sub-isr.data} we compare the ISR data with other
measurements, in particular with the recent results of the ZEUS
collaboration for inclusive neutron production in the
photoabsorption reaction $\gamma p\to Xn$. The two sets of data turn
out to be not really consistent, what makes questionable the
normalization of the ISR data.

We summarize the main results and observations in Sect.~\ref{summary}.

\section{Pion pole}\label{pion.pole}

\begin{figure}[htb]
\centerline{
  \scalebox{0.45}{\includegraphics{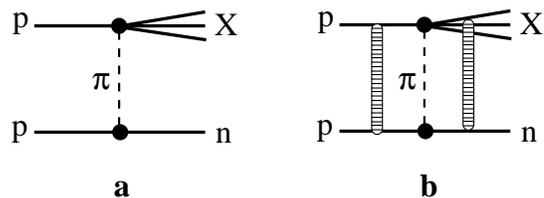}}
 }
\caption{\label{pion}
 {\bf a:} Born graph with single pion exchange;
{\bf b:} illustration of absorptive corrections.
 }
 \end{figure}

The Born approximation pion exchange contribution to the amplitude
of neutron production $pp\to nX$, depicted in Fig.~\ref{pion}a, in
the leading order in small parameter $m_N/\sqrt{s}$ has the form
 \beq
A^B_{p\to n}(\vec q,z)=\frac{1}{\sqrt{z}}\,
\bar\xi_n\left[\sigma_3\,\tilde q_L+
\vec\sigma\cdot\vec q_T\right]\xi_p\,
\phi^B(q_T,z)\,,
\label{100}
 \eeq
 where $\vec\sigma$ are Pauli matrices;  $\xi_{p,n}$ are the proton or
neutron spinors;  $\vec q_T$ is the transverse component of the momentum transfer;
 \beq
\tilde q_L=(1-z)\,m_N\,.
\label{110}
 \eeq

In the region of small $1-z\ll1$ the pseudoscalar amplitude $\phi^B(q_T,z)$ has
the
triple-Regge form,
 \beqn
\phi^B(q_T,z)&=&\frac{\alpha_\pi^\prime}{8}\,
G_{\pi^+pn}(t)\,\eta_\pi(t)\,
(1-z)^{-\alpha_\pi(t)}
\nonumber\\ &\times&
A_{\pi^+ p\to X}(M_X^2)\,,
\label{120}
 \eeqn
 where the 4-momentum
transfer squared $t$ has the form,
 \beq
t=-\,{1\over z}\,\left(\tilde q_L^2+q_T^2\right)\,,
\label{130}
 \eeq
 and $\eta_\pi(t)$ is the phase (signature) factor which can be expanded near
the pion pole as,
 \beq
\eta_\pi(t)=i-ctg\left[\frac{\pi\alpha_\pi(t)}{2}\right]\approx
i+\frac{2}{\pi\alpha_\pi^\prime}\,
\frac{1}{m_\pi^2-t}\,.
\label{140}
 \eeq
 We assume a linear pion Regge trajectory
$\alpha_\pi(t)=\alpha_\pi^\prime(t-m_\pi^2)$,
where $\alpha_\pi^\prime\approx 0.9\GeV^{-2}$.
The imaginary part in (\ref{140}) is neglected in what follows, since its 
contribution near the pion pole is small.

The effective vertex function
$G_{\pi^+pn}(t)=g_{\pi^+pn}\exp(R_1^2t)$ includes the pion-nucleon
coupling and the form factor which incorporates the $t$-dependence
of the coupling and of the $\pi N$ inelastic amplitude. We take the
values of the parameters used in \cite{k2p},
$g^2_{\pi^+pn}(t)/8\pi=13.85$ and $R_1^2=0.3\GeV^{-2}$.
Notice that the choice of $R_1$ does not bring much uncertainty, since we focus here at 
data for forward production, $q_T=0$, so $t$ is quite small.

The amplitudes in (\ref{100})-(\ref{120}) are normalized as,
 \beq
\sigma^{\pi^+ p}_{tot}(s'=M_X^2)={1\over M_X^2}
\sum\limits_X|A_{\pi^+ p\to X}(M_X^2)|^2\,,
\label{144}
 \eeq
 where different hadronic final states $X$ are summed at fixed invariant
mass $M_X$. Correspondingly, the differential cross section of inclusive
neutron production reads \cite{bishari,2klp},
 \beqn
z\,\frac{d\sigma^B_{p\to n}}{dz\,dq_T^2}&=&{1\over s}
\left|A^B_{p\to n}(\vec q_T,z)\right|^2
\nonumber\\ &=&
\left(\frac{\alpha_\pi^\prime}{8}\right)^2
|t|G_{\pi^+pn}^2(t)\left|\eta_\pi(t)\right|^2
(1-z)^{1-2\alpha_\pi(t)}
\nonumber\\ &\times&
\sigma^{\pi^+ p}_{tot}(s'=M_X^2)\,.
\label{146}
 \eeqn

Since at $z\to1$ the value of $M_X^2$ decreases, we rely on a
realistic fit to the experimental data \cite{pdg} for $\pi^+p$ total
cross section.

\begin{figure}[htb]
\centerline{
  \scalebox{0.5}{\includegraphics{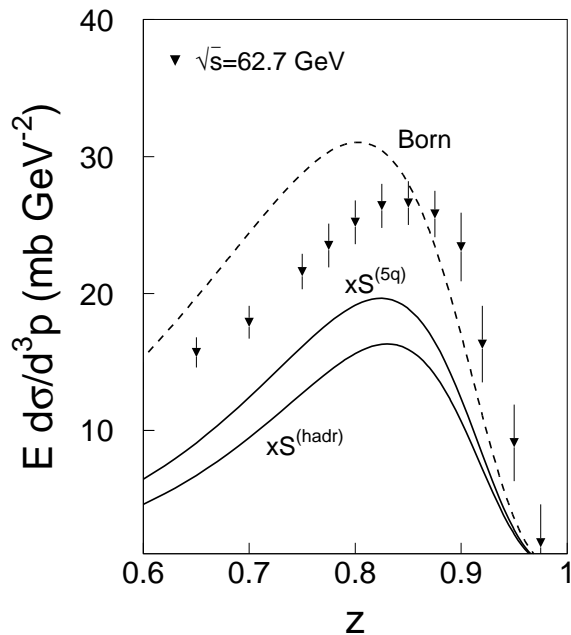}}}
\caption{\label{fig:isr1}
 Born approximation (dashed curve) for leading neutron production and ISR
data \cite{isr}, at $\sqrt{s}=62.7\GeV$ and $p_T=0$. Two solid
curves, the upper and bottom ones, show the effect of absorptive
corrections calculated in the dipole approach ($\times S^{(5q)}$)
and in hadronic representation ($\times S^{(hadr)}$) respectively.}
 \end{figure}

\begin{figure}[htb]
\centerline{
  \scalebox{0.5}{\includegraphics{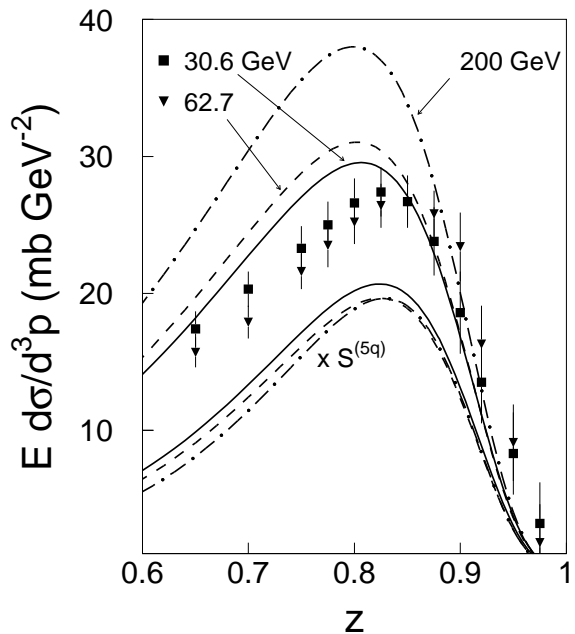}}
 }
\caption{\label{fig:isr2}
 Energy dependence of inclusive neutron production. The three
upper curves present the forward cross section at
$\sqrt{s}=30.6\GeV$ (solid), $62.7\GeV$ (dashed) and $200\GeV$
(dotted-dashed), calculated in the Born approximation. The same
cross sections, although corrected for absorption ($\times
S^{(5q)}$), are given by the three curves at the bottom. Data at
$\sqrt{s}=30.6\GeV$ and $62.7\GeV$ \cite{isr} are depicted by
squares and inverse triangles respectively}
 \end{figure}

 The results of the Born approximation calculation, Eq.~(\ref{146}), at
$\sqrt{s}=200,\ 62.7$ and $30.6\GeV$, are depicted together with the
ISR data \cite{isr}, in Figs.~\ref{fig:isr1} and \ref{fig:isr2}.

 The data are given at two energies $\sqrt{s}=30.6\GeV$ and
 $62.7\GeV$, and therefore we use these energies in our calculations. One can see
that the Born approximation considerably exceeds the data.

Notice that only at small $1-z \sim m_\pi/m_N$ one can approach the
pion pole, i.e. the smallness of the pion mass is important for
Eq.~(\ref{120}). Otherwise $t$ is large even at $q_T=0$, and the
pion exchange gains a considerable imaginary part. Besides, the
spin-flip amplitude $\phi^B(q_T,z)$ acquires a weak dependence on
$q_T$ at small scattering angles, $q_T^2\ll(1-z)^2m_N^2$.

\section{Absorptive corrections}\label{absorptive.corrections}

Absorptive corrections, or initial/final state interactions,
illustrated in Fig.~\ref{pion}, look quite complicated in momentum
representation where they require multi-loop integrations. However,
if they do not correlate with the amplitude of the process
$\pi^+p\to X$, then these corrections factorize in impact parameter
and become much simpler. Therefore, first of all, we should Fourier
transform the amplitude Eq.~(\ref{100}) to impact parameter space.

\subsection{Impact parameter representation}\label{sub-impact}

The partial Born amplitude at impact parameter $\vec b$,
corresponding to (\ref{100}), has the form,
 \beq
f^B_{p\to n}(\vec b,z)=\frac{1}{\sqrt{z}}\,
\bar\xi_n\left[\sigma_3\,\tilde q_L\,\theta^B_0(b,z)-
i\,\frac{\vec\sigma\cdot\vec b}{b}\,
\theta^B_s(b,z)\right]\xi_p\,,
\label{150}
 \eeq
 where
\beqn
\theta^B_0(b,z) &=& \int d^2q\,e^{i\vec b\vec q}\,
\phi^B(q_T,z)
\nonumber\\ &=&
\frac{N(z)}{1-\beta^2\epsilon^2}\,
\left[K_0(\epsilon b)-K_0(b/\beta)\right]\,;
\label{154}
 \eeqn

 \beqn
\theta^B_s(b,z) &=& {1\over b}
\int d^2q\,e^{i\vec b\vec q}\,
(\vec b\cdot\vec q)\,\phi^B(q_T,z)
\nonumber\\ &=&
\frac{N(z)}{1-\beta^2\epsilon^2}\,
\left[\epsilon\,K_1(\epsilon b)-\frac{1}{\beta}\,K_1(b/\beta)\right]\,.
\label{164}
 \eeqn
 Here
 \beqn
N(z) &=&\frac{1}{2}\,g_{\pi^+pn}\,
z(1-z)^{\alpha^\prime_\pi(m_\pi^2+\tilde q_L^2/z)}
e^{-R_1^2 \tilde q_L^2/z}
\nonumber\\&\times&
A_{\pi p\to X}(M_X^2)\,
\nonumber\\
\epsilon^2&=&\tilde q_L^2+zm_\pi^2\,,
\nonumber\\
\beta^2&=&R_1^2-\alpha_\pi^\prime\,\frac{\ln(1-z)}{z}
\,.
\label{166}
 \eeqn
 To simplify the calculations we replaced here the Gaussian form factor,
$\exp(-\beta^2q_T^2)$, by the monopole form $1/(1+\beta^2q_T^2)$,
which is a good approximation at the small values of $q_T$ we are
interested in. At the same time we keep the exact expression for the
dependence on $\tilde q_L$, which can be rather large.

\subsection{Survival amplitude of large rapidity gaps}\label{sub-survival}

At large $z\to1$ the process under consideration is associated with
the creation of a large rapidity gap (LRG), $\Delta y=|\ln(1-z)|$,
where no particle is produced. Absorptive corrections may also be
interpreted as a suppression related to the survival probability of
LRG, which otherwise can be easily filled by multiparticle
production initiated by inelastic interactions of the projectile
partons with the target. Usually the corrected cross section is
calculated as a convolution of the cross section with the survival
probability factor (see \cite{ryskin} and references therein). This
recipe may work sometimes as an approximation, but only for
$q_T$-integrated cross section. Otherwise one should rely on a
survival amplitude, rather than probability. Besides, the absorptive
corrections should be calculated differently for the spin-flip and
non-flip amplitudes (see below).

In impact parameter representation one can expand the incoming proton over
the Fock components, $|3q\ra,\ |3qg\ra,\ |4q\bar q\ra$, etc. For every
Fock state with fixed transverse separations between the constituents
the eikonal form is exact. In the dipole representation the absorption
corrected amplitude can be written as,
 \beqn
f_{p\to n}(b,z) &=& \sum\limits_l
\prod\limits_{i} d^2r_i\,d\alpha_i\,
C^p_l(\{r_i,\alpha_i\})\,
\nonumber\\ &\times&
\left[\tilde f^B_{p\to n}(b,z,\{r_i,\alpha_i\})\right]_l
e^{if_l(b,z,\{r_i\})}.
\label{170}
 \eeqn
 Here we sum over Fock states containing different number of partons of
different species, having transverse positions $\vec r_i$ and
fractional light-cone momenta $\alpha_i$. The parton distribution
amplitudes $C^p_l(\{r_i,\alpha_i\})$ are normalized to the
probabilities $W_l$ of having $l$-th Fock state in the proton, $\int
\prod\limits_i d^2r_i\,d\alpha_i |C^p_l(\{r_i,\alpha_i\})|^2=W_l$.
We neglect the small real part of the partial amplitude
$f_l(b,z,\{r_i\})$ of elastic scattering of the partonic state
$|l;\{r_i\}\ra$ on a nucleon, and assume that it is pure imaginary
and isotopic invariant (Pomeron exchange).

\begin{figure}[htb]
\centerline{
  \scalebox{0.32}{\includegraphics{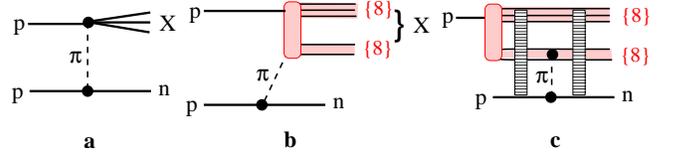}}
 }
\caption{\label{fock-pp}
 {\bf a:} Born graph with single pion exchange and excitation of the projectile
proton, $p+\pi\to X$;
 {\bf b:} inelastic proton-pion interaction, $p+\pi\to X$, via
color exchange, leading to the production of two color-octet dipoles
which hadronize further to $X$;
 {\bf c:} Fock state representation of the previous mechanism. A color
octet-octet dipole which is a 5-quark Fock component of the projectile proton,
interacts with the target proton via
$\pi^+$ exchange. This 5-quark state may experience initial and final state
interaction via vacuum quantum number (Pomeron) exchange with the nucleons
(ladder-like strips). }
 \end{figure}

Now we have to identify the Fock states responsible for initial and
final state interactions leading to absorptive corrections. We start
with Fig.~\ref{fock-pp}a, containing the amplitude of the
pion-proton inelastic collision $\pi+p\to X$. This is usually
described as color exchange, leading to the creation of two color
octet states with a large rapidity interval $\sim \ln(M_X^2/s_0)$
($s_0=1\GeV^2$), as illustrated in Fig.~\ref{fock-pp}b.
 Perturbatively, the interaction is mediated by gluonic exchanges.
Nonperturbatively, e.g. in the string model, the hadron collision
looks like intersection and flip of strings. Hadronization of the
color-octet dipole (described for example by the string model) leads
to the production of different final states $X$.

According to Fig.~\ref{fock-pp}b the produced color octet-octet
state can experience final state interactions with the recoil
neutron. On the other hand, at high energies multiple interactions
become coherent, and one cannot specify at which point the
charge-exchange interaction happens, i.e. both initial and final
state interactions must be included. One can rephrase this in terms
of the Fock state decomposition. The projectile proton can fluctuate
into a 5-quark color octet-octet before the interaction with the
target. The fluctuation life-time, or coherence time (length), is
given by
 \beq
l_c=\frac{2E_p}{M_X^2-m_N^2}\,,
\label{190}
 \eeq
 which rises with energy and at high energies considerably exceeds the
longitudinal size of target proton. Technically, one should integrate the
amplitude over the longitudinal coordinate $l$ of the fluctuation point,
weighted with a phase factor $e^{il/l_c}$ (see an example in \cite{kst2}),
which effectively restricts the distances from the target to $\Delta
l\lsim l_c$.

This leads to a different space-time picture of the process at high
energies, namely the incoming proton fluctuates into a 5-quark state
$|\{3q\}_8\{\bar qq\}_8\ra$ long in advance of the interaction between the
$\{\bar qq\}_8$ pair and the target via pion exchange, see
Fig.~\ref{fock-pp}c. This is the general intuitive picture which is
supported by more formal calculations \cite{zamolodchikov,stan-ivan}.
Assuming only final state interactions one should sum up the amplitudes of
the process depicted in Fig.~\ref{fock-pp}b and of the double step
collision in which the 5-quark state is produced diffractively in the
first collision $pN\to |\{3q\}_8\{\bar qq\}_8\ra\,N$, and then the 5-quark
system experiences charge exchange scattering of another proton via pion
exchange. The resulting amplitude exposes both initial and final state
attenuation of the 5-quark state,
 \beq
f_{p\to n}(b,z)=f^B_{p\to n}(b,z)\,S^(b,z)\,.
\label{195}
 \eeq

Thus, the 5-quark component of the projectile proton propagates through
the target experiencing initial and final state interactions. The
effective absorption cross section is the inelastic cross section of the
$|\{3q\}_8\{\bar qq\}_8\ra$ dipole on a nucleon.

Of course, besides the five valence quarks, also gluons can be radiated,
which are essential for the energy dependence of $\sigma^{\pi
p}_{tot}(M_X^2)$. They are effectively included in the following
calculations.

\subsection{Reggeon calculus}\label{sub-reggeons}

\begin{figure}[htb]
\centerline{
  \scalebox{0.42}{\includegraphics{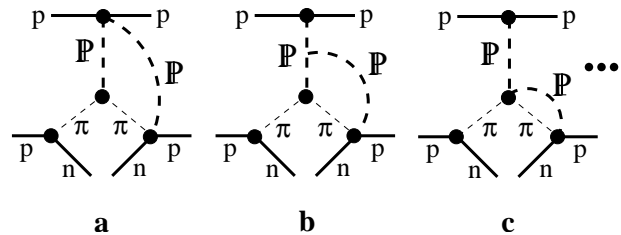}}
 } \caption{ Absorptive corrections due to possibility of inelastic
interactions which can fill up the large rapidity gap. {\bf a:} Interactions
of the projectile proton and its remnants (see Fig.~\ref{vertex}) with the
target ; {\bf b:} triple Pomeron interaction due to interactions of produced
particles (e.g. radiated gluons); {\bf c:} interactions including the pion
remnants (see Fig.~\ref{vertex}). Only some of the interference graphs are
shown.}
 \label{graphs}
 \end{figure}

Previous calculations \cite{3n,ryskin} proposed rather mild absorptive
corrections, corresponding to only a beam proton experiencing multiple
interactions in the target. This was motivated by Reggeon graphs depicted
in Fig.~\ref{graphs}a,b (we show only some of the interference terms))

\begin{figure}[htb]
\centerline{
  \scalebox{0.45}{\includegraphics{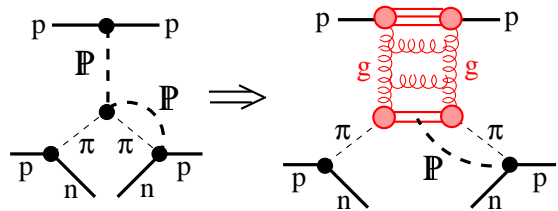}}
 }
\caption{Structure of the four-Reggeon vertex $\pi\pi\Pom\Pom$.}
\label{vertex}
 \end{figure}

 Fig.~\ref{graphs}a presents multiple interactions of the projectile
proton and its remnants. Fig.~\ref{graphs}b includes interactions of the
multiparton states produced in $\pi-p$ inelastic collision (see
Fig.~\ref{pion}). This term is proportional to the triple-Pomeron coupling,
which is assumed to be small, and for this reason it was neglected in
\cite{3n,ap,ryskin}. The third term Fig.~\ref{graphs}c, overlooked in
\cite{3n,ap}, has a different behavior\footnote{This graph was considered in
\cite{ryskin}, but without detailed analysis.} since it contains a 4-Reggeon
vertex $\pi\pi\Pom\Pom$, and may not be small. The structure of this
vertex, as well as of the cut Pomeron, are shown in Fig.~\ref{vertex}.

 The interaction of the radiated gluons (the rungs of the Pomeron ladder) is
indeed weak, as follows from the smallness of the triple-Pomeron coupling. This
is explained dynamically in \cite{spots} by the shortness of the transverse
separation between the radiated gluons and the source. There is no such a
suppression, however, for the interaction of the ${\bar qq}_8$ pair, which is the
pion remnant, as is depicted in Fig.~\ref{vertex}. Calculations performed
below confirm that the term shown in Figs.~\ref{graphs}c, \ref{vertex}, missed
in \cite{3n,ryskin}, is large.

\section{Absorptive corrections in saturated regime}\label{dipole}

\begin{figure}[htb]
\centerline{
  \scalebox{0.35}{\includegraphics{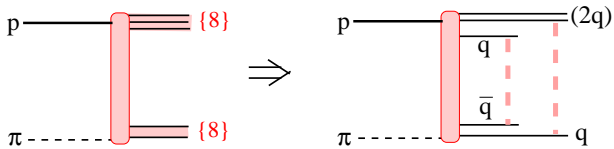}}
 }
\caption{Inelastic pion-proton interaction, $\pi+p\to X$,
in Fig.~\ref{fock-pp}, leading to the production of two color-triplet dipoles,
$q-\bar q$ and $(2q)-q$.
 }
\label{dual}
 \end{figure}

Another way to estimate the absorption effects is to consider directly
the interaction of the 5-quark octet-octet dipole with the proton target.
Following  the dual parton model \cite{capella} approach, we replace the
$|3q\ra_8-|\bar qq\ra_8$ dipole by two color triplet dipoles, $(qq)-q$
and $q-\bar q$, as is illustrated in Fig.~\ref{dual}.
 This approximation has an accuracy $1/N_c^2$, which is sufficient for our
purposes.

Thus, the survival amplitude for such a 5-quark state can be represented
as a product,
 \beqn
S^{(5q)}(b) &=& S^{(3q)}(b)\,S^{(q\bar q)}(b)
\label{290}
\\ &=&
\left[1-\Im\Gamma^{(3q)p}(b)\right]
\left[1-\Im\Gamma^{(\bar qq)p}(b)\right].
 \nonumber
 \eeqn
 similar to Eq.~(\ref{240}),
The elastic amplitude $\Gamma^{(\bar33)p}(b)$ of a color $\{\bar 33\}$
dipole interacting with a proton is related to the partial elastic amplitude
 \beq
\Im\Gamma^{(\bar 33)p}(b,z)=
\int d^2r W_{\bar 33}(r,M_X^2)\,
\Im f^{\bar33}_{el}(\vec b,\vec r,s,\alpha),
\label{295}
 \eeq
 where $\alpha$ is the fractional light-cone momentum carried by the $3$, or $\bar
3$;  $r$ is the dipole transverse size;  and $W_{\bar 33}(r,M_X^2)$ is the
dipole size distribution function, which is specified later, as well as the relation
between $\alpha$ and $z$. Now we concentrate on the partial dipole amplitude
$f^{\bar33}_{el}(\vec b,\vec r,s,\alpha)$.

\subsection{Generalized unintegrated gluon density and partial dipole
amplitude}\label{sub-partial.amplitude}

The $\bar qq$-dipole-proton total cross section can be directly fitted to data
on the proton structure function measured in deep-inelastic scattering (DIS).
The popular form \cite{gbw} of the dipole cross section, which describes quite well
data at small Bjorken $x$, has a saturated shape, i.e. the cross section levels
off at large dipole sizes. For soft reactions, such as the one we are dealing with here, the c.m.
energy rather than Bjorken $x$, is the proper variable. A similar
parameterization, with the saturated shape fitted to data on DIS at
$Q^2$ not high and real photo-absorption and photoproduction of vector mesons, led to
the result \cite{kst2},
 \beq
\sigma_{\bar qq}(r,s)=\sigma_0(s)\left[1-e^{-r^2/R_0^2(s)}\right]\,,
\label{300}
 \eeq
 where $R_0(s)=0.88\,fm\,(s_0/s)^{0.14}$ and $s_0=1000\,GeV^2$. This cross
section is normalized to reproduce the pion-proton total cross section, $\int
d^2r\,|\Psi_\pi(r)|^2\sigma_{\bar qq}(r,s)=\sigma^{\pi p}_{tot}(s)$.  The pion
wave function squared integrated over longitudinal quark momenta has the
form,
 \beq
\left|\Psi_\pi(\vec r)\right|^2 =
\frac{3}{8\pi \la r^2_{ch}\ra_\pi}
\exp\left(-\frac{3r^2}{8\la r^2_{ch}\ra_\pi}\right)\,,
\label{310}
 \eeq
 where $\la r^2_{ch}\ra_\pi=0.44\fm^2$ \cite{r-pion} is the mean pion charge
radius squared. This normalization condition results in
 \beq
\sigma_0(s)=\sigma^{\pi p}_{tot}(s)\,
\left(1 + \frac{3\,R^2_0(s)}{8\,\la r^2_{ch}\ra_{\pi}}
\right)\,,
\label{320}
 \eeq
 For the numerical calculation we rely on one of the popular parameterizations for the
energy dependent total cross sections \cite{pdg} (only the Pomeron
part), $\sigma^{\pi p}_{tot}(s)=\Sigma_0+\Sigma_1\ln^2(s/s_1)$,
where $\Sigma_0=20.9\mb$ $\Sigma_1=0.31\mb$ and $s_1=28.9\GeV^2$.

Just as the dipole-proton total cross section can be calculated via the
unintegrated gluon density in the proton \cite{gbw}, one can calculate the
partial amplitude $f(\vec b,\vec r)$ via a generalized transversely
off-diagonal gluon distribution \cite{amir2},
 \beqn
\im f^N_{\bar qq}(\vec b,\vec r,\beta)
&=&\frac{1}{12\pi}
\int\frac{d^2q\,d^2q'}{q^2\,q'^2}\,\alpha_s\,
{\cal F}(x,\vec q,\vec q^{\,\prime})
e^{i\vec b\cdot(\vec q-\vec q^{\,\prime})}
\nonumber\\ &\times&
\left(e^{-i\vec q\cdot\vec r\beta}-
e^{i\vec q\cdot\vec r(1-\beta)}\right)\,
\nonumber\\ &\times&
\left(e^{i\vec q^{\,\prime}\cdot\vec r\beta}-
e^{-i\vec q^{\,\prime}\cdot\vec r(1-\beta)}\right)\,
\,.
\label{325}
 \eeqn
 A model for the generalized unintegrated gluon density was proposed
recently \cite{amir2}, based on the saturated form of the diagonal
gluon density \cite{gbw}, and assuming a factorized dependence on
both $\vec q$ and $\vec q^{\,\prime}$. One gets
 \beqn
&& {\cal F}(x,\vec q,\vec q^{\ \prime}) =
\frac{3\,\sigma_0}{16\,\pi^2\,\alpha_s}\ q^2\,q'^{\,2}\,R_0^2(x)
\nonumber\\ &\times&
{\rm exp}\Bigl[-{1\over8}\,R_0^2(x)\,(q^2+q'^{\,2})\Bigr]
{\rm exp}\bigl[-{1\over2}B(x)(\vec q-\vec q^{\ \prime})^2\bigr]
\,,\nonumber\\
 \label{330}
 \eeqn
 This Bjorken $x$-dependent density, appropriate for hard reactions, leads to
an $x$-dependent partial amplitude \cite{amir2}. Although in general
it should not be used for soft processes, one can switch from an
$x$- to an $s$-dependence keeping the same parameterization and
adjusting the parameters to observables in soft reactions, as was
done in \cite{kst2}, see Eq.~(\ref{300}). Then the partial amplitude
reads
\begin{widetext}

 \beqn
\Im f^{\bar qq}_{el}(\vec b,\vec r,s,\alpha) =
\frac{\sigma_0(s)}{8\pi B(s)}\,
\Biggl\{\exp\left[-\frac{[\vec b+\vec r(1-\alpha)]^2}{2B(s)}\right] +
\exp\left[-\frac{(\vec b-\vec r\alpha)^2}{2B(s)}\right]-
2\exp\Biggl[-\frac{r^2}{R_0^2(s)}
-\frac{[\vec b+(1/2-\alpha)\vec r]^2}{2B(s)}\Biggr]
\Biggr\},\nonumber\\
\label{340}
 \eeqn
\end{widetext}
 This partial amplitude correctly reproduces the dipole
cross section
Eq.~(\ref{300}),
 \beq
2\int d^2b\,{\rm Im}f^{\bar qq}_{el}(\vec b,\vec r,s,\alpha)=
\sigma_{\bar qq}(r,s)\,.
\label{360}
 \eeq

Another condition that needs to be satisfied is reproducing the
slope $B^{\pi p}_{el}(s)$ of the elastic $\pi p$ differential cross
section,
 \beqn
 B^{\pi p}_{el}(s)&=&{1\over 2}\,\la b^2\ra
\frac{1}{\sigma^{\pi p}_{tot}}
\int d^2b\int\limits_0^1 d\alpha
\nonumber\\ &\times&
\int d^2r\,
\left|\Psi_\pi(\vec r,\alpha)\right|^2
{\rm Im}f^{\bar qq}_{el}(\vec b,\vec r,s,\alpha)
\,.
\label{380}
 \eeqn
 This condition allows to evaluate the parameter $B(s)$ in (\ref{340}). To
simplify this calculation, we fix here $\alpha=1/2$ in the partial
amplitude and arrive at
 \beq
B(s)=B^{\pi p}_{el}(s)-{1\over3}\,\la r_{ch}^2\ra_\pi
- {1\over8}\,R_0^2(s)\,.
\label{400}
 \eeq
 In what follows we use a Regge parameterization for the elastic slope, $B^{\pi
p}_{el}(s)=B_0+2\alpha_\Pom^\prime \ln(s/\mu^2)$, with $B_0=6\GeV^{-2}$,
$\alpha_\Pom^\prime=0.25\GeV^{-2}$, and $\mu^2=1\GeV^2$.

In the case of a $(2q)-q$ dipole all relations are analogous to
Eqs.~(\ref{300})-(\ref{400}), but one should make the following replacements:
(i) $\sigma^{\pi p}_{tot}(s)\Rightarrow\sigma^{pp}_{tot}(s)$ with
$\Sigma_0=35.5\mb$; (ii) $\la r_{ch}^2\ra_\pi\Rightarrow\la
r_{ch}^2\ra_p=0.8\fm^2$ \cite{proton}; (iii) $B^{\pi p}_{el}(s)\Rightarrow
B^{pp}_{el}(s)$ with $B_0=8\GeV^{-2}$.

\subsection{Survival amplitudes of dipoles}\label{sub-dipole.survival}

To proceed further with the calculation of the survival amplitude,
Eqs.~(\ref{290})-(\ref{295}), we have to specify the dipole size
distribution. One can get a hint from Figs.~\ref{fock-pp}b and \ref{dual}
that the size distribution of the $(3q)_8-(\bar qq)_8$ dipoles is actually
given by the partial amplitude squared of $\pi-p$ elastic scattering at
c.m. energy $E_{c.m.}=M_X=\sqrt{s(1-z)}$. Assuming a Gaussian dependence of
this partial amplitude on impact parameter, we get
 \beq
W_{8-8}(r,M_X^2)= \frac{1}{2\pi\,B^{\pi p}_{el}(M_X^2)}
\exp\left[-\frac{r^2}{2B^{\pi p}_{el}(M_X^2)}\right]\,.
\label{420}
 \eeq
 Thus, the size of the $q\bar q$ and $q-2q$ dipoles is $z$-dependent and
controlled by $B^{\pi p}_{el}(M_X^2)$.

 Performing the integration in (\ref{295}) with this weight factor and the partial
dipole amplitude Eq.~(\ref{340}), we arrive at the survival amplitude for a
$\bar q-q$ dipole,
 \begin{widetext}
\beqn
S^{(\bar qq)}(b,z) &=& 1-\frac{\sigma_0(s)}{4\pi}
\left\{\frac{1}{B_\alpha(s,z)}
\exp\left[-\frac{b^2}
{B_\alpha(s,z)}\right] +
\frac{1}{B_{1-\alpha}(s,z)}
\exp\left[-\frac{b^2}
{B_{1-\alpha}(s,z)}\right]
\right.\nonumber\\ &-& \left.
\frac{2}
{B_{1/2-\alpha}(s,z)
\left[1+B^{\pi p}_{el}(M_X^2)/R_0^2(s)\right]}
\exp\left[-\frac{b^2}
{B_{1/2-\alpha}(s,z)}\right]
\right\}\,,
\label{440}
\eeqn
 \end{widetext}
 where
 \beqn
B_\beta(s,z) &=& 2B(s)+\beta^2\,B^{\pi p}_{el}(M_X^2)\,,
\label{460}
 \eeqn
 and $\beta$ equals either  $\alpha$, or $1-\alpha$, or $1/2-\alpha$.
All other quantities related to a $\bar qq$ dipole are defined in
Sect.~\ref{sub-partial.amplitude}.

The same expressions Eqs.~(\ref{440})-(\ref{460}) can be used for the
survival amplitude $S^{(3q)}(b)$ of a baryon $(2q)-q$ dipole, after making
the same replacements of $\sigma^{\pi p}_{tot}(s)$, $\la r_{ch}^2\ra_\pi$
and $B^{\pi p}_{el}(s)$, as is listed at the end of
Sect.~\ref{sub-partial.amplitude} (except $\tilde B^{\pi p}_{el}(M_X^2)$
which should be kept as is).

The last variable to be specified is $\alpha$, which is related to
$z=1-M_X^2/s$ via the relation for the invariant mass $M_X$ of the
5q system,
 \beq
M_X^2=\frac{m^2_{3q}+k_T^2}{1-\alpha}+
\frac{m^2_{\bar qq}+k_T^2}{\alpha}\,,
\label{470}
 \eeq
 where $k_T$ is the relative transverse momentum of $(\bar qq)_8$ and $(3q)_8$.
For the large values of $M_X^2\gg m_p^2$ that we are interested in,
 \beq
\alpha= \frac{m_T^2}{M_X^2}=\frac{m_T^2}{s(1-z)}\,,
\label{480}
 \eeq
 where we fix $m_T^2=\la m^2_{\bar qq}+k_T^2\ra=1\GeV^2$, assuming that
$\la m^2_{\bar qq}\ra\sim \la k_T^2\ra \sim m_\rho^2$.

\begin{figure}[htb]
\centerline{
  \scalebox{0.45}{\includegraphics{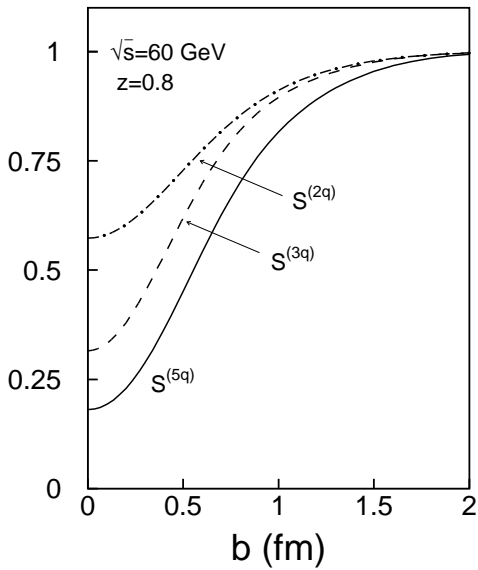}}}
 \caption{ Partial survival amplitude $S(b,z)$ at $\sqrt{s}=60\GeV$ and
$z=0.8$. Survival amplitudes $S^{(2q})(b,z)$ for a $\bar q-q$
dipole, and $S^{(3q})(b,z)$ for a $q-2q$ dipole, are depicted by
dot-dashed and dashed curves, respectively. Their product,
$S^{(5q})(b,z)$, is shown by the solid curve. }
 \label{survival1}
 \end{figure}

\begin{figure}[htb]
\centerline{
  \scalebox{0.45}{\includegraphics{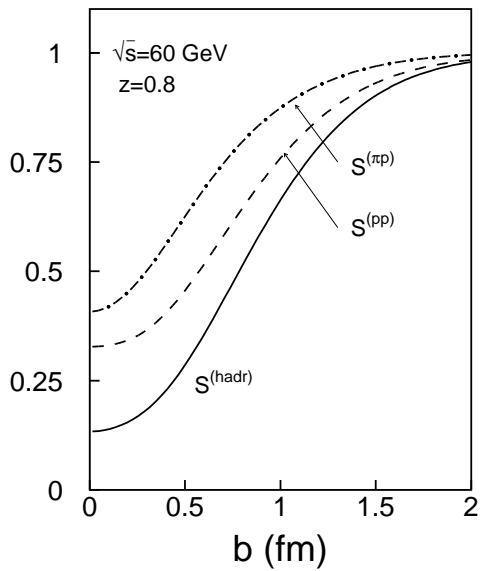}}
}
 \caption{ Partial survival amplitude $S(b,z)$ at $\sqrt{s}=60\GeV$ and
$z=0.8$.  The survival amplitude evaluated in hadronic
representation. Dot-dashed, dashed and solid curves show the pion
and proton survival amplitudes and their product, respectively.}
 \label{survival2}
 \end{figure}

The results for the $5q$ dipole survival probability Eq.~(\ref{290})
calculated at $\sqrt{s}=44.7\GeV$ and $z=0.8$, are shown in
Figs.~\ref{survival1} and \ref{survival2}.

\section{Survival amplitude in hadronic representation}\label{hadronic}

\subsection{Expansion over multi-hadronic states}\label{sub-multihadrons}

One can expand the 5-quark Fock state over the hadronic basis,
 \beq
\left|\{3q\}_8\{\bar qq\}_8\right\ra =
d_0|p\ra + d_1|N\pi\ra + d_2|N2\pi\ra + ...\,.
\label{200}
 \eeq

These components are associated with different suppression factors,
which can be calculated via known hadron-proton elastic amplitudes.
Correspondingly, the absorption corrected partial amplitude gets the
form
 \beq
f_{p\to n}(b,z)=f^B_{p\to n}(b,z)\,S^{(hadr)}(b)\,,
\label{220}
 \eeq
 where $S^{(hadr)}(b)$ is the survival amplitude averaged over different
hadronic components in (\ref{200}).

Since the admixture of sea quarks in the proton is small, the
projection of the 5-quark state to the proton, the amplitude $d_0$,
must be small. The states that contribute consist mainly of a
nucleon accompanied by one or more pions and other mesons, and
therefore here we make the natural assumption that the amplitude
$d_1$ is the dominant one, since both states $|\{3q\}_8\{\bar
qq\}_8\ra$ and $|N\pi\ra$ have the same valence quark content.
Then the survival amplitude of a large rapidity gap mediated by pion
exchange is related to the amplitude of no-interaction of a $p-\pi$
pair propagating through the target proton.  Neglecting the
difference in impact parameters of the pion and proton, we get
 \beqn
S^{(hadr)}(b) &=& S^{\pi p}(b)\,S^{pp}(b)
\nonumber\\ &=&
\left[1-\Im\Gamma^{pp}(b)\right]
\left[1-\Im\Gamma^{\pi p}(b)\right]\,.
\label{240}
 \eeqn
 Here we expressed the hadron-nucleon survival amplitude via the elastic
partial amplitude $\Gamma(b)$,
 \beq
S^{hN}(b)=1-\Im\Gamma^{hN}(b)\,.
\label{210}
 \eeq
 An implicit energy dependence is assumed in here and further on,
unless specified.

Nevertheless, the calculation of the partial amplitudes
$\Gamma^{hN}(b)$ is still a challenge, and different models and
approximations are known. For instance, if the total cross section
$\sigma^{hN}_{tot}$ and the elastic slope $B^{hN}_{el}$ are known,
and one assumes a Gaussian shape for the differential hadron-proton
cross section, one gets
 \beq
\Im\Gamma^{hN}_{(Gauss)}(b)=
\frac{\sigma^{hN}_{tot}}
{4\pi B^{hN}_{el}}\,
\exp\left[-\frac{b^2}{2B^{hN}_{el}}\right]\,.
\label{260}
 \eeq
 At high energies, however, this is a poor approximation, since the
unitarity bound stops the rise of the partial amplitude at small
$b$, and the periphery becomes the main source of the observed rise
of the total cross sections \cite{amaldi,k3p}. As a result, the
shape of the $b$-dependence changes with energy and cannot be
Gaussian.

One has to incorporate unitarity corrections, and a popular way to do it
is the eikonal approximation \cite{kps1},
 \beq
\Im\Gamma^{hp}_{(eik)}(b) =1 -
e^{-{\rm Im}\Gamma_0^{hp}(b)}\,,
\label{280}
 \eeq
 where $\Gamma_0^{hp}(b)$ is an input, bare amplitude, which is actually
unknown. It can be compared with data only after unitarization (e.g.
eikonalization) procedure.

The eikonal approximation cannot be correct, since hadrons are not
eigenstates of the interaction, and they can be diffractively
excited. To improve the eikonal approximation (\ref{280}) one should
include all possible intermediate diffractive excitations
\cite{gribov}. This is a difficult task, since there is no
experimental information about diffractive off-diagonal transitions
between different excited states. So far this has been done only in
a two-channel toy-model \cite{kl-78,levin}.

Another way of include the higher order Gribov corrections is the so
called quasi-eikonal model \cite{kaidalov}. However, it is based on
an ad hoc recipe for higher order diffractive terms, which is not
supported by any known dynamics.

The dipole approach \cite{zkl,mine,kps1} allows to sum up the Gribov
corrections in all orders, for a given Fock state of the projectile
hadron. However, the inclusion of higher Fock states is difficult
and model dependent.

\subsection{Partial elastic amplitude from data}\label{sub-data}

Nevertheless, one can get reliable information about
$\Gamma^{hp}(b)$ extracting it directly from data for the elastic
differential cross section and the ratio of real-to-imaginary
amplitudes.  We parameterize the imaginary and real parts of the
elastic scattering amplitude in momentum representation as
  \beq
{\rm Im}\,f^{hp}(t)=\sum\limits_{i=1}^{3}
a_i\,e^{b_i\,t};
\label{data.1}
 \eeq
 \beq
{\rm Re}\,f^{hp}(t)=c\,e^{d\,t}
\label{data.2}\ ,
 \eeq
 where $a_i,\ b_i,\ c,\ d$ are the fitting parameters. The
amplitudes are related to the cross sections as
 \beq
\frac{d\,\sigma^{hp}_{el}}{d\,t}=
\bigl[{\rm Re}\,f^{hp}(t)\bigr]^2 +
\bigl[{\rm Im}\,f^{hp}(t)\bigr]^2\ ;
\label{data.3}
 \eeq
 \beq
\sigma^{hp}_{tot}=4\,\sqrt{\pi}\,{\rm Im}\,f^{hp}(0)\ .
\label{data.4}
 \eeq

We applied this analysis to data on the $pp$ elastic differential
cross section \cite{data-pp}. To make the normalization of data for
the differential cross section more certain, first of all we perform
a common fit of the $pp$ and $\bar pp$ total cross sections with the
same Pomeron part, as function of energy.  Then we adjust the
normalizations of data for the differential elastic cross sections
to the optical points, {\it i.e.} demand that $4\,\sqrt{\pi}\,\sum
a_i=\sigma_{tot}$ at each energy.

Data \cite{rho} for the ratio of real to imaginary parts of the
forward amplitude, $\rho^{hp}(s)={\rm Re}\,f^{hp}(0)/{\rm
Im}\,f^{hp}(0)$, were also used in the analysis.  We fitted these
data with a smooth energy dependence and demanded $c=\rho\,\sum a_i$
for each energy included in the analysis of differential cross
sections. The details of the fit to $pp$ data can be found in
\cite{k3p}. Here we applied the same procedure to data for
pion-proton scattering, using the database from \cite{data-ppi}.

After the parameters in (\ref{data.1}) and (\ref{data.2}) are found,
one can calculate the partial amplitude in impact parameter
representation at each energy as
 \beq
\Gamma^{hp}(b)=\frac{1}{2\,\pi^{3/2}}
\int d^2b\,e^{i\,\vec q\cdot\vec b}\,f^{hp}(-q^2)\ ,
\label{data.5}
\eeq
 where $\vec q$ is the transverse component of the transferred momentum,
$t\approx - q^2$. It is normalized according to (\ref{data.4}).

\begin{figure}[htb]
\centerline{
  \scalebox{0.3}{\includegraphics{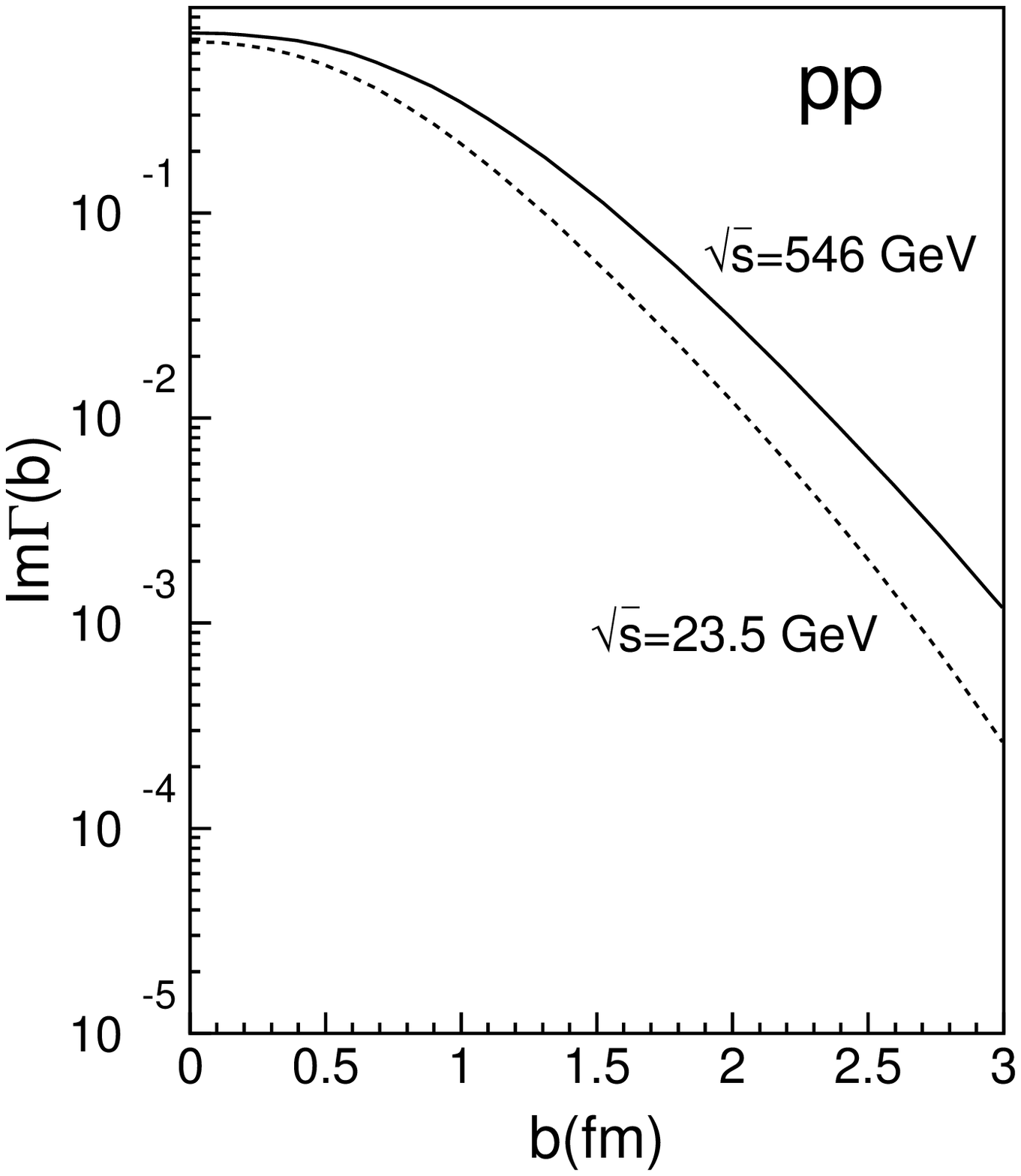}}
  \scalebox{0.3}{\includegraphics{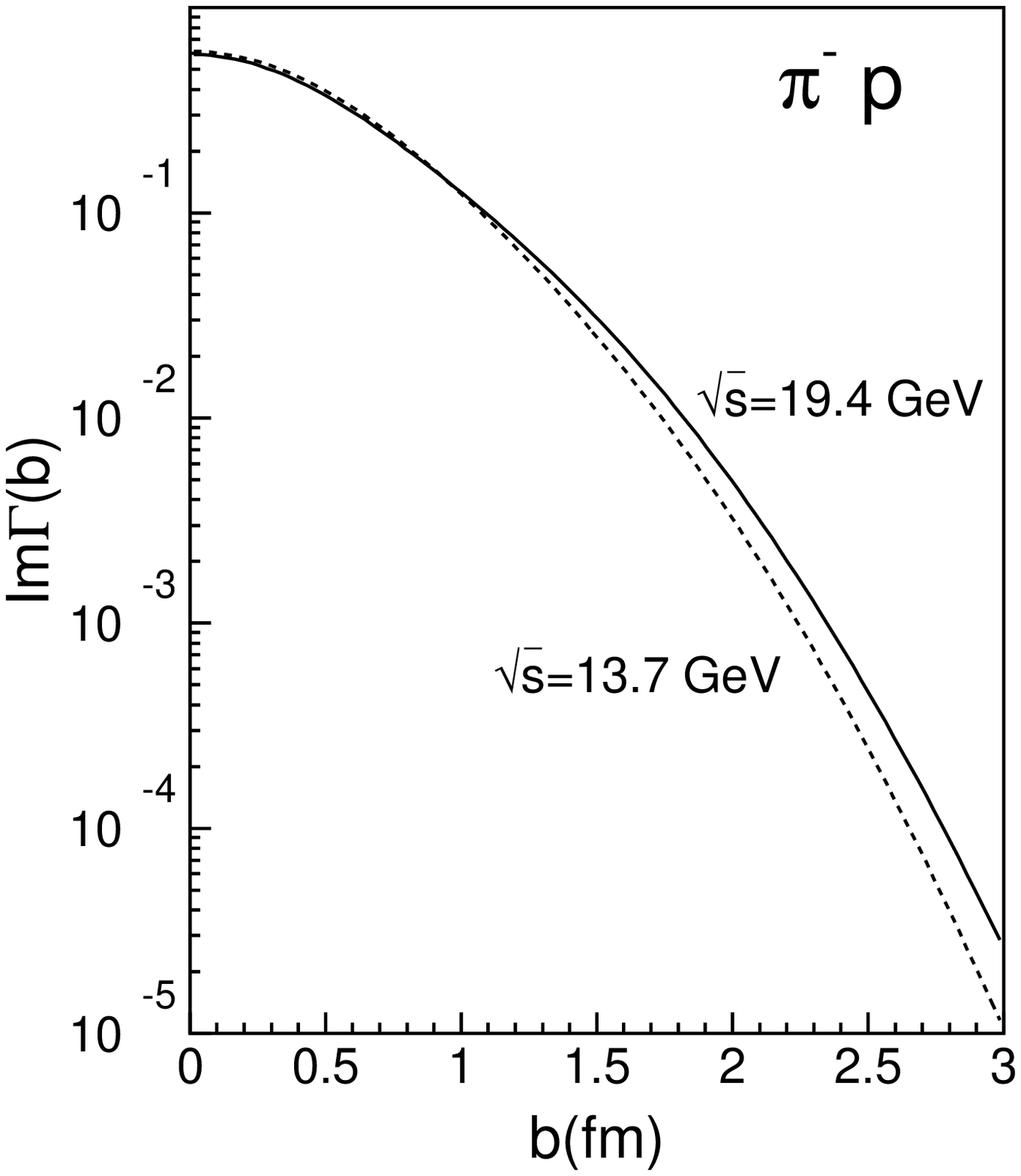}} }
 \caption{ Imaginary part of the partial elastic amplitude extracted by a
model-independent analysis of data on the elastic differential cross
section. {\it Left:} $pp$ partial amplitude ${\rm
Im}\,\Gamma^{pp}(b)$ at c.m. energies $\sqrt{s}=23.5\GeV$ and
$546\GeV$. {\it Right:} $\Im\Gamma^{\pi p}(b)$ at
$\sqrt{s}=13.7\GeV$ and $19.4\GeV$.}
 \label{partial}
 \end{figure}

 Examples are depicted in Fig.~\ref{partial} for the partial amplitudes ${\rm
Im}\,\Gamma^{pp}(b)$ (left panel) and ${\rm Im}\,\Gamma^{\pi p}(b)$ (right
panel).

One can see that at $b=0$ the amplitude nearly saturates the unitarity
limit and hardly changes with energy, while at larger impact parameters
the amplitude substantially grows. This means that the corresponding
LRG survival amplitude is minimal for central collisions where it
steadily decreases with energy towards zero in the black disc (Froissart)
limit. Our results for $S^{(hadr)}(b,z)$ are depicted in
Fig.~\ref{survival2} at $\sqrt{s}=40\GeV$ and $z=0.8,\ 0.9$.

\subsection{Extreme damping}\label{extreme}

Although the survival amplitudes for protons and pions were
extracted in a model independent way directly from data, we feel
that the main assumption made above, that the 5-quark state can be
represented by just a $\pi N$ pair has a rather shaky basis. Quite
probably the higher Fock component containing more pions might be
important. Indeed, either the color octet-octet state or the two
triplet-antitriplets representing its decay multiply produce
hadrons, mainly pions. Of course, it would be exaggeration to
include all of these pions into the absorption damping factor. This
would be like interpreting the color transparency effect in hadronic
representation by a sum of different hadrons. Neglecting the off
diagonal transitions and interferences one arrives at the so called
Bjorken paradox \cite{bjorken}: instead of color transparency one
gets hadronic opacity. The most economic way to include the
interferences is to switch to the color dipole representation, as we
did in Sect.~\ref{dipole}. However, it useful to understand the
magnitude of a maximal suppression when all produced pions
contribute in the same footing to the absorption corrections.

Apparently the pion multiplicity should rise with $M_X^2$. Following
the prescription of the dual parton model \cite{capella} we replaced
the octet-octet dipole, $\{3q\}_8 - \{\bar qq\}_8$, by two
color-triplet strings, $q-\bar q$ and $qq-q$, which share the c.m.
energy $M_X$ in fractions of $1/3$ and $2/3$ respectively. This is
illustrated in Fig.~\ref{dual}.

The multiplicities of pions produced from the decay of these strings
are known from fits to data on $e^+e^-$ annihilation \cite{tasso}
and deep-inelastic scattering \cite{kaidalov-ter},
 \beqn
\la n_\pi\ra_{q-\bar q}&=&
4+0.72\ln(M_X^2/9s_0)\,;
\label{202}\\
\la n_\pi\ra_{qq-q}&=&
0.45+0.135\ln(4M_X^2/9s_0)\,,
\label{204}
 \eeqn
 where $s_0=1\GeV$. Since we need the full multiplicity, we multiplied the
number of charged pions by $3/2$. The fit Eq.~(\ref{202}) was performed
for $M_X>4.2\GeV$, which, for instance at $\sqrt{s}=50\GeV$, corresponds
to $z<0.99$. We impose this restriction which is well within the interval
of $z$ we are interested in.

Thus we can replace the $|\{3q\}_8\{\bar qq\}_8\ra$ dipole by a
nucleon and multipion state. In the eikonal approach such a maximal
suppression corresponds to the absorptive suppression factor,
 \beq
S^{(hadr)}_{max}(b,z)=S^{NN}(b)
\sum\limits_{n_\pi=0} W_{n_\pi}(z)\,S^{(n_\pi\pi)N}(b)\,,
\label{222}
 \eeq
 where $W_{n_\pi}(z)$ is the probability distribution of number of pions
which we assume to have a Poisson shape, $W_{n_\pi}(z)=(\la
n_\pi\ra^{n_\pi}/n_\pi!)e^{-\la n_\pi\ra}$. The mean number of pions
$\la n_\pi(z)\ra$ depends on $z$ according to
(\ref{202})-(\ref{204}) and equals to,
 \beqn
\la n_\pi(z)\ra&=&\la n_\pi\ra_{q-\bar q}+\la n_\pi\ra_{qq-q}
\nonumber\\&=&
2.76+0.855\,\ln(M_X^2/s_0)\,.
\label{206}
 \eeqn

The survival amplitude of a LRG for the target nucleon interacting with a
row of pions can be presented in the eikonal form like in the Glauber
model, i.e. $S^{(n_\pi\pi)N}(b)=[S^{\pi N}(b)]^{n_\pi}$. Then the
maximal suppression factor Eq.~(\ref{222}) gets the form,
 \beqn
&&S^{(hadr)}_{max}(b,z)=S^{NN}(b)
\exp\left\{-\la n_\pi(z)\ra\left[1-S^{\pi N}(b)\right]\right\}
\nonumber\\ &=&
\left[1-\Im\Gamma^{NN}(b)\right]
\exp\Bigl[-\la n_\pi(z)\ra\,
\Im\Gamma^{\pi N}(b)\Bigr]\,.
\label{208}
 \eeqn
 Later, in Sect.~\ref{discussion} we will compare the effect of the maximal
suppression Eq.~(\ref{208}) with the conventional ones.

\section{Cross section corrected for absorption}\label{xsection}

\begin{figure}[htb]
\centerline{
  \scalebox{0.45}{\includegraphics{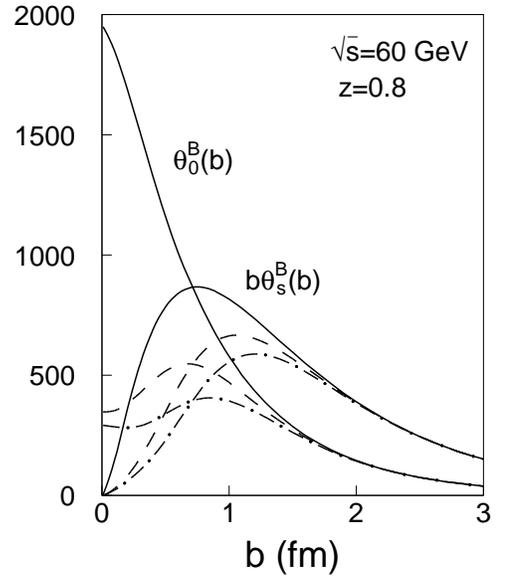}}
}
 \caption{Partial spin amplitudes, Eq.~(\ref{150}), for neutron production,
non-flip, $\theta_0(b,z)$, and spin-flip, $b\theta_s(b,z)$. Solid curves show
the result of Born approximation. Dashed and dot-dashed curves include
absorptive corrections calculated in the dipole approach ($\times
S^{(5q)}(b,z)$) and in hadronic model ($\times S^{(hadr)}(b,z)$),
respectively.}
 \label{spin-amplitudes}
 \end{figure}

Now we can correct for absorption the Born partial amplitudes Eq.~(\ref{150})
of neutron production,
 \beq
\theta_{0,s}(b,z)=
\theta^B_{0,s}(b,z)\,
S(b,z)\,,
\label{500}
 \eeq
 where $S(b,z)$ is calculated either within the dipole approach,
Eq.~(\ref{290}), or in the hadronic model, Eq.~(\ref{240}). In
Fig.~\ref{spin-amplitudes} we compare the Born partial spin
amplitudes with the ones corrected for absorption, plotted as
functions of impact parameter at $z=0.8$ and $\sqrt{s}=44.7\GeV$.

 Now, it is straightforward to Fourier transform these amplitudes back to
momentum representation. The absorption modified Eq.~(\ref{100})
reads
 \beq
A_{p\to n}(\vec q,z)=\frac{1}{\sqrt{z}}
\bar\xi_n\left[\sigma_3 \tilde q_L\,\phi_0(q_T,z)+
\vec\sigma\cdot\vec q_T\phi_s(q_T,z)\right]\xi_p,
\label{520}
 \eeq
 where according to (\ref{154}), (\ref{164}) and (\ref{220}),
 \beqn
\phi_0(q_T,z)&=&\frac{N(z)}{2\pi(1-\beta^2\epsilon^2)}
\int\limits_0^\infty db\,b\,J_0(bq_T)\,
S(b,z)
\nonumber\\ &\times&
\left[K_0(\epsilon b)-K_0\left({b\over\beta}\right)\right]
\,;
\label{540}
 \eeqn

 \beqn
q_T\,\phi_s(q_T,z)&=&\frac{N(z)}{2\pi(1-\beta^2\epsilon^2)}
\int\limits_0^\infty db\,b\,J_1(bq_T)\,
S(b,z)
\nonumber\\ &\times&
\left[\epsilon\, K_1(\epsilon b)-
{1\over\beta}\,K_1\left({b\over\beta}\right)\right]
\,.
\label{560}
 \eeqn

Eventually, we are in a position to calculate the differential cross
section of inclusive production of neutrons corrected for absorption,
\beq
z\,\frac{d\sigma_{p\to n}}{dz\,dq_T^2}=
\sigma_0(z,q_T) + \sigma_s(z,q_T)\,,
\label{580}
 \eeq
 where
 \beqn
\sigma_0(z,q_T)&=&
\frac{\tilde q_L^2}{zs}\,
\left|\phi_0(q_T,z)\right|^2
\label{590}\\
\sigma_s(z,q_T)&=&
\frac{q_T^2}{zs}\,
\left|\phi_s(q_T,z)\right|^2\,.
\label{600}
 \eeqn
 The forward neutron production cross section corrected for absorption is
compared with data \cite{isr} in Fig.~\ref{fig:isr1}. The two models
for absorption, dipole and hadronic, give the upper and bottom solid
curves respectively. The results of both models are pretty close to
each other, but substantially underestimate the data (see further
discussions). This is a consequence of very strong absorptive
corrections found here compared to previous calculations
\cite{3n,ap}, which nevertheless reported good agreement with data.

The energy dependence of the cross section is presented in
Fig.~\ref{fig:isr2}, at $\sqrt{s}=30.6,\ 62.7$ and $200\GeV$.
Apparently the steep rise of the cross section with energy, observed
in Born approximation, is nearly compensated by the falling energy
dependence of the LRG survival amplitudes. Aside for the
normalization, the results for the $z$- and energy-dependence agree
quite well with the data.

\begin{figure}[htb]
\centerline{
  \scalebox{0.45}{\includegraphics{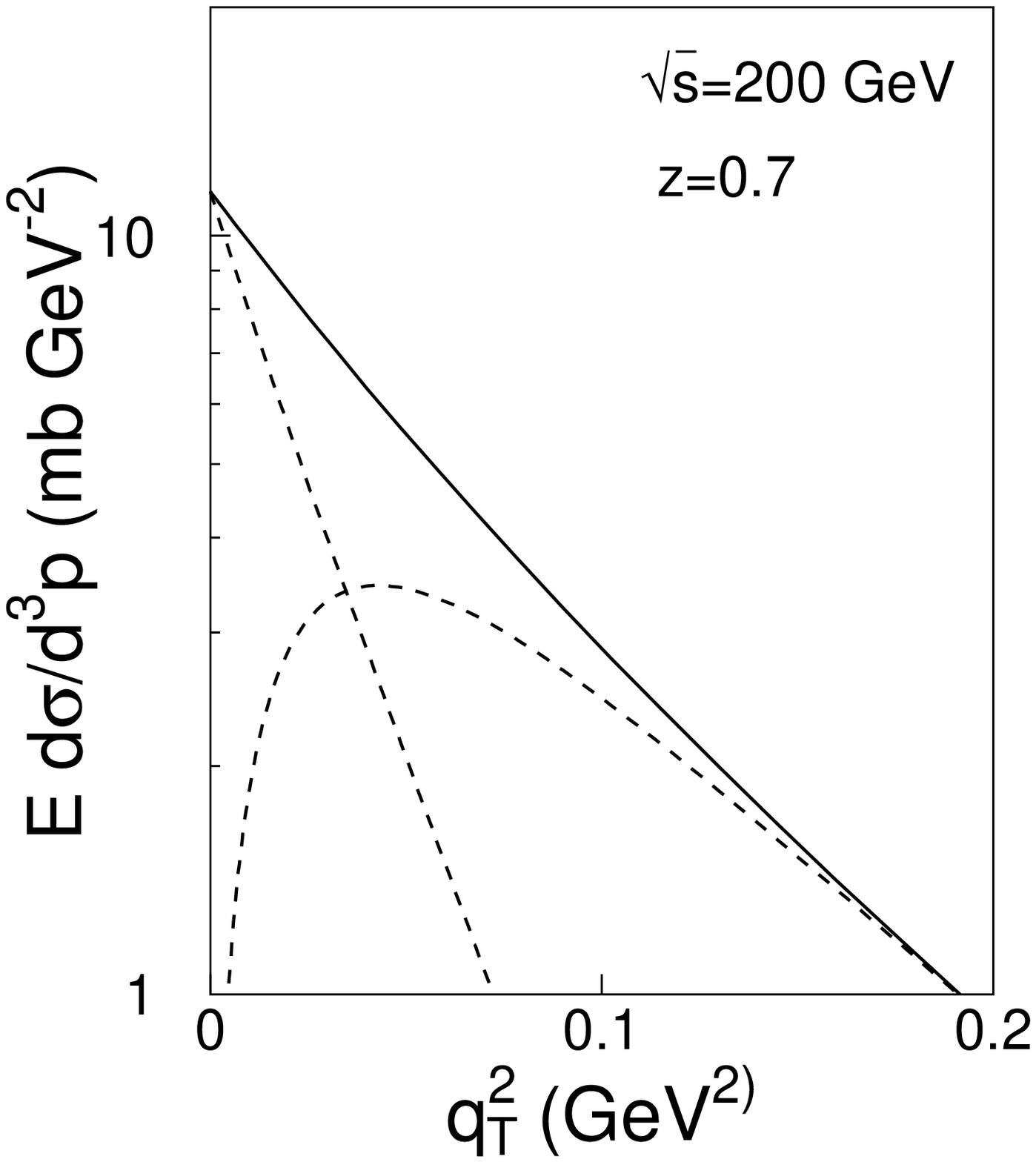}}}
\centerline{
  \scalebox{0.45}{\includegraphics{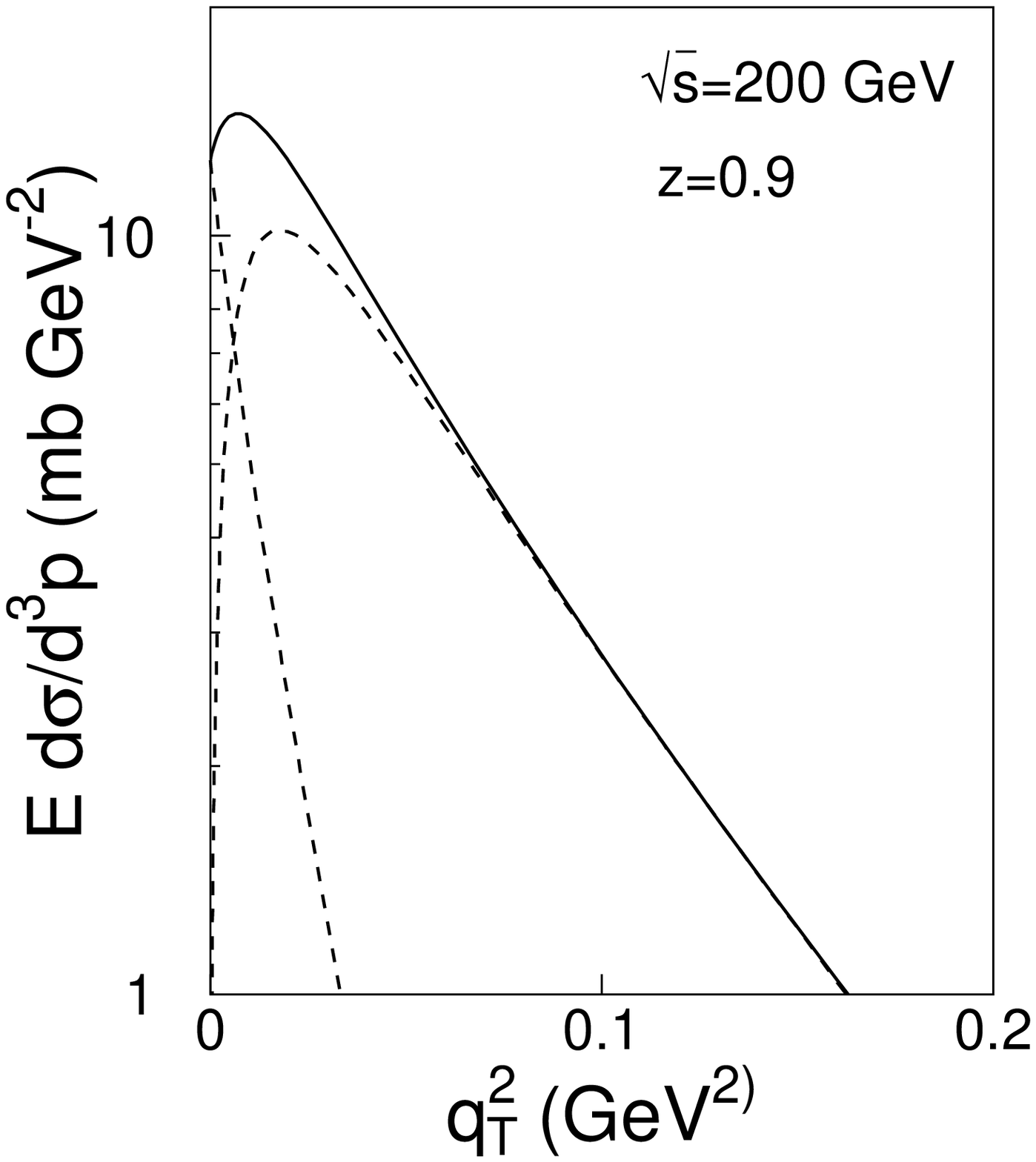}}
}
 \caption{Differential cross section of neutron production,
Eq.~(\ref{580}), at $\sqrt{s}=200\GeV$, $z=0.7$ (upper panel) and $z=0.9$
(bottom panel).  Contributions of the non-flip, Eq.~(\ref{590}), and
spin-flip, Eq.~(\ref{600}), processes are shown by dashed curves, and
their sum is depicted by solid curves.}
 \label{sigma-qt}
 \end{figure}

We also calculate the $q_T$-dependence of the differential cross section
Eq.~(\ref{580}). The results for $\sqrt{s}=200\GeV$ are shown in
Fig.~\ref{sigma-qt} for $z=0.6$ (left panel)  and $z=0.9$ (right panel).
 The $q_T$ distribution shrinks towards larger $z$. For instance, the slope
calculated at $q_T^2=0.1\GeV^2$ equals to $B(z=0.7)=12.3\GeV^{-2}$
and $B(z=0.9)=17.3\GeV^{-2}$. At the same time, at small $q_T$ the
spin-flip term starts sticking out at large $z$, and the effective
slope measured at such small $q_T$ may become small, and even
negative.

Notice that the effective slope also rises with energy. The $q_T$
distribution calculated at $\sqrt{s}=50\GeV$ at the same values of $z$
demonstrates a similar pattern, but the slopes are about two units of
$\GeV^{-2}$ smaller.

\section{Discussion}\label{discussion}

There are few points in the above presentation which deserve more
discussion.

\subsection{Maximal suppression}

\begin{figure}[htb]
\centerline{
  \scalebox{0.4}{\includegraphics{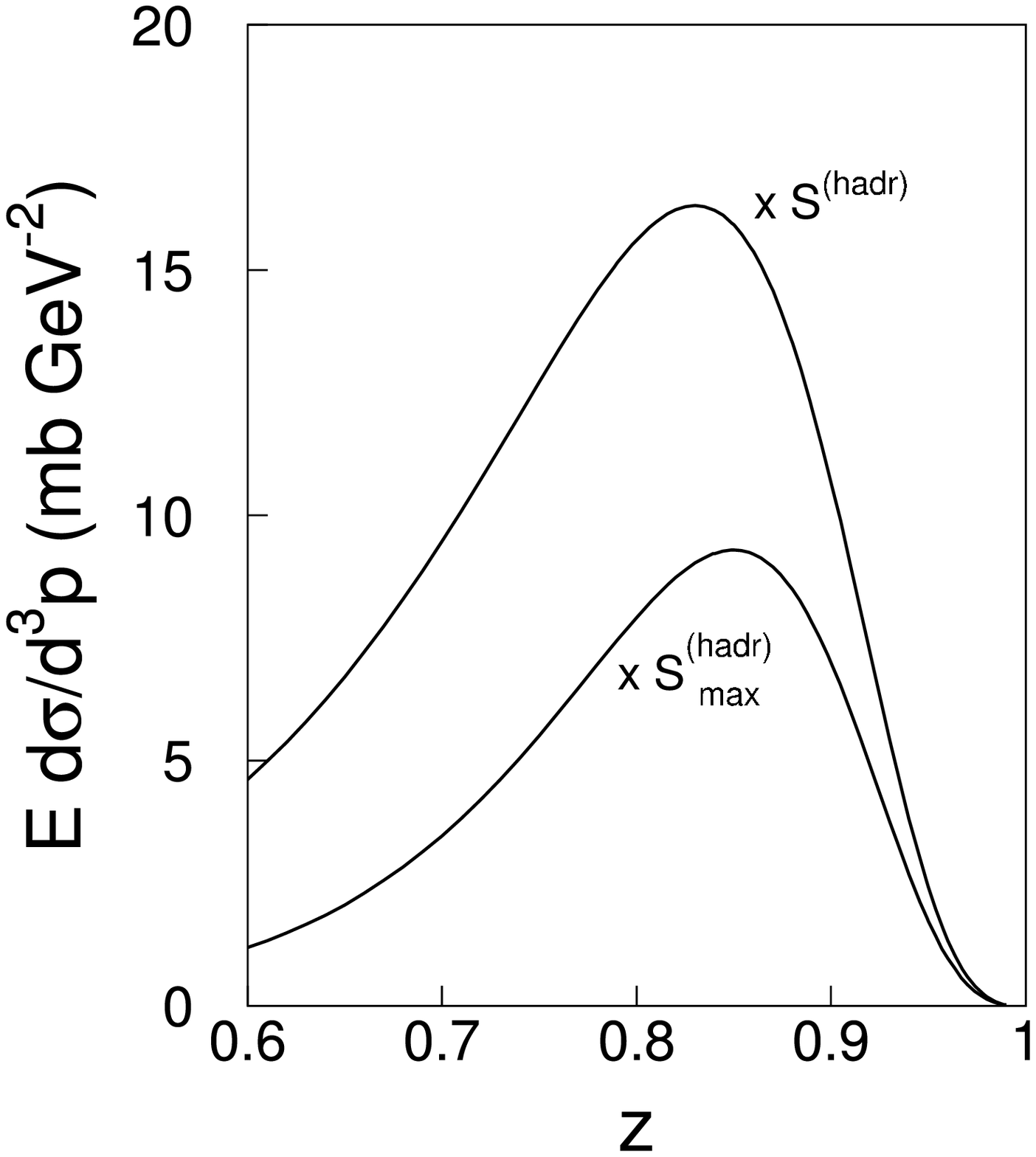}}
}
 \caption{Comparison of the effect of the cross section damping caused by a $\pi-p$ 
pair and by a nucleon accompanied by $\la n_\pi\ra$ pions (see Sect.~\ref{extreme} 
for the details), represented by the upper and bottom curves respectively.
Calculations are performed for $\sqrt{s}=30\GeV$ and $q_T=0$.}
 \label{max}
 \end{figure}

Although our results presented in Figs.~\ref{fig:isr1} and \ref{fig:isr2} for the 
cross section calculated with the hadronic model are quite below the ISR data, we 
think that we could only underestimate the strength of the absorptive damping. We 
represented the color octet-octet dipole by by a $\pi p $ pair, but apparently the 
effective number of pions might be larger. Of course this can only suppress the cross 
section further down and worsen the disagreement with the ISR data. To see the scale 
of possible effects we considered in Sect.~\ref{extreme} an extreme case of mean 
number of pions corresponding to hadronization of the octet-octet dipole. The result 
for the cross section of neutron production is compared with the $\pi p$ hadronic 
model in Fig.~\ref{max}.
 The effect of suppression caused by the extra pions is not strong at large $z$,
since the pion exchange partial amplitude is very peripheral, while the suppression 
factor $S^{(hadr)}_{max}$ is more central. Correspondingly, the effect of extra 
suppression becomes stronger towards smaller $z$.

 \subsection{Challenging the ISR data}\label{sub-isr.data}

 The shape of both the $z$ and energy dependence which resulted from our
calculations agree with data \cite{isr}. However, the predicted
cross section, shown in Figs.~\ref{fig:isr1} and \ref{fig:isr2},
underestimates the data \cite{isr} by about a factor of two.

\begin{figure}[htb]
\centerline{
  \scalebox{0.45}{\includegraphics{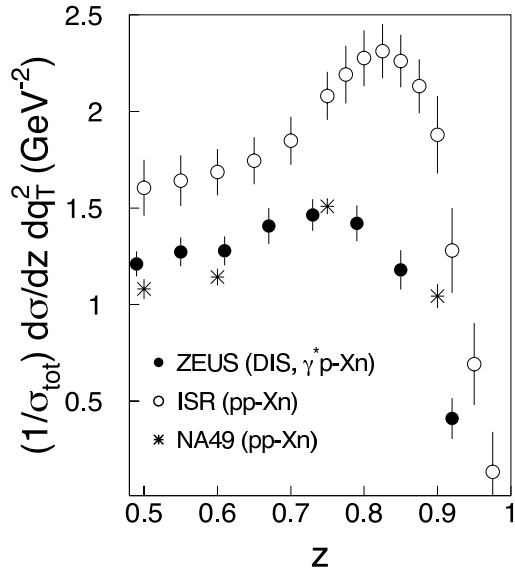}}
}
 \caption{Number of events distribution, Eq.~(\ref{620}), for neutron
production. {\it Open points:} ISR data \cite{isr} for forward,
$q_T=0$, neutron production divided by $\sigma^{pp}_{tot}$ at
$\sqrt{s}=62.7\GeV$ \cite{pdg}. The overall normalization
uncertainty is $20\%$ \cite{isr}. {\it Closed points:} number of
events for neutron production in DIS ($Q^2>4\GeV^2$). The ZEUS data
\cite{zeus} are extrapolated to $q_T=0$ as is described in the text.
Systematic errors related to the acceptance and energy scale
uncertainties are added in quadrature.  The overall normalization
uncertainty is $4\%$ \cite{zeus}. {\it Asterix points:} event number distribution for $pp\to nX$ 
measured in the NA49 experiment at $E_{lab}=158\GeV$ and extrapolated to $q_T=0$ \cite{na49}.
}
 \label{isr-zeus}
 \end{figure}

Nevertheless, there are indications that the source of disagreement
may be the normalization of the data. A strong evidence comes from
the recent measurements by the ZEUS collaboration \cite{zeus} of
leading neutron production in semi-inclusive deep-inelastic
scattering (DIS) and photoproduction, that the normalization of the
ISR data \cite{isr} is overestimated by about a factor of two.
Indeed, according to Regge factorization the fraction of events with
leading neutron production in $h$-proton collision,
 \beq
\frac{dN}{dz dq_T^2}=
\frac{1}{\sigma^{hp}_{tot}}\,
\frac{d\sigma_{hp\to Xn}}{dz dq_T^2}\,,
\label{620}
 \eeq
 should be universal, i.e. independent of the particle $h$.
Of course this universality should be broken by absorption
corrections, and it is natural to expect that neutron damping should
be stronger in $pp$ collisions than in photoproduction. However, a
comparison of photo-production and $pp$ data performed in
\cite{zeus} demonstrated just the opposite: the ratio
Eq.~(\ref{620}) for $pp$ is twice that for photoproduction.
Moreover, Fig.~\ref{isr-zeus} demonstrates that even neutrons
produced in DIS, where absorption effects should be minimal, are
quite more suppressed than in the ISR data for $pp$ collisions.
 Extrapolating to $q_T=0$ the ZEUS data for neutron production in DIS, within
an angle $0.8\,$mrad, we used the measured slope $b(z)=(16.3\,z -
4.25)\GeV^{-2}$.

 Notice that the ZEUS results \cite{zeus} also show that the ratio Eq.~({620})
rises with $Q^2$, demonstrating decreasing absorptive corrections,
in good accord with the above expectations and in contradiction with
the weak absorption suggested by the ISR data.

Another evidence comes from the ratio of the pion-to-proton
structure functions measured at small $x$ in \cite{zeus}. Contrary
to the natural expectation $F_2^\pi(x)/F_2^p(x)\approx 2/3$, it was
found to be about $1/3$. This shows that the absorptive corrections
reduce the cross section by a factor of two (like in our
calculations). As was already commented, absorptive corrections in
$pp$ collisions should not be smaller than in DIS.

Although the systematic uncertainty of the ISR data was claimed in
\cite{isr} to be $20\%$, it was probably underestimated.

One can find in \cite{ryskin} more comments on the current
controversies in the available data for leading neutron production
in hadronic collisions.

A firm support for our conjecture about an incorrect normalization of the ISR data comes from preliminary data from the NA49 experiment at CERN SPS\cite{na49} for leading neutron production in $pp$ collisions at $E_{lab}=158\GeV$. The measured cross section integrated over $q_T$ was extrapolated to $q_T=0$ assuming the same slope of $q_T$ dependence as measured for proton production \cite{na49}. The found fractional cross section plotted in  Fig.~\ref{isr-zeus} is about twice as low as the ISR data, but agrees well with the ZEUS DIS data.

\subsection{Further corrections}\label{sub-firther.corrections}

Besides the pion pole, Fig.~\ref{pion}, other mechanisms which were
discussed in \cite{k2p} can contribute. Isovector Reggeons, $\rho$
$a_2$ and $a_1$, also lead to neutron production. These Reggeons
contribute mostly to the spin-flip amplitude, i.e. vanish in the
forward direction where we compare with data. These corrections to
the cross section were estimated in \cite{k2p} to be about $10\%$,
as well as the possibility of additional pion production in the
pion-nucleon vertex, $\pi p\to \pi n$ \cite{k2p}. We neglect this
corrections here, since they are small and quite uncertain. The main
focus of this paper is the calculation of absorptive corrections.

Since the isovector Reggeon amplitudes are mainly spin-flip, they are small in forward direction, but become more important with rising $q_T$. Thus,
they should reduce the value of the $q_T^2$-slope of the differential cross section calculated in Sect.~\ref{xsection}. Indeed the slope measured in the ZEUS experiment \cite{zeus2007} is substantially smaller than is suggested by the  contribution of pion exchange.

\section{Summary}\label{summary}

To summarize, we highlight some of the results.
\begin{itemize}

\item Pion exchange is usually associated with the spin-flip amplitude.
However, the amplitude of an inclusive process mediated by pion
exchange acquires a substantial non-flip part which in many cases
dominates.

\item We applied absorptive corrections to the spin amplitudes.
This is quite different from a convolution of the LRG survival
probability with the cross section, as it has been done in many
publications. We found that the non-flip amplitude is suppressed by
absorption much more than the spin-flip one, therefore applying an
overall suppression factor is not correct.

\item
We identified the projectile system which undergoes initial and
final state interactions as a color octet-octet 5-quark state.
Absorptive corrections are calculated within two models,
color-dipole light-cone approach, and in hadronic representation.
The two descriptions, being so different, nevertheless lead to very
similar results.

\item Since the projectile 5-quark state interacts with the target
stronger than a single nucleon, we predict a much stronger damping of
neutrons compared to some of previous estimates.

\item
Comparison of fractional cross sections of forward neutron
production in $pp$ collisions \cite{isr} and in DIS \cite{zeus} show
a substantial discrepancy which indicates an incorrect normalization
of ISR data. The preliminary data for neutron production in $pp$ collisions
from the NA49 experiment at CERN SPS \cite{na49} are about twice lower than 
the ISR data, once again confirming that the latter has an incorrect normalization.
This explains why our results are significantly lower
than the ISR data. 
New data for inclusive neutron production at RHIC, at
$\sqrt{s}=200-500\GeV$ are expected soon  \cite{phenix}.

\end{itemize}

\begin{acknowledgments}

We are grateful to Misha Ryskin
for informative discussions, and to Hans Gerhard Fischer and Dezso Varga
for providing us with preliminary data from the NA49 experiment and for useful comments. This work was supported in part
by Fondecyt (Chile) grants 1050519 and 1050589, and by DFG (Germany)
grant PI182/3-1.
\end{acknowledgments}

\end{document}